# Hippocampal and striatal involvement in cognitive tasks: a computational model

Fabian Chersi [1*], Neil Burgess [1]

**ABSTRACT**

The hippocampus and the striatum support episodic and procedural memory, respectively, and "place" and "response" learning within spatial navigation. Recently this dichotomy has been linked to "model-based" and "model-free" reinforcement learning.

Here we present a well-constrained neural model of how both systems support spatial navigation, and apply the same model to more abstract problems such as sequential decision making. In particular, we show that if a task can be transformed into a Markov Decision Process, the machinery provided by the hippocampus and striatum can be utilized to solve it.

These results show how the hippocampal complex can represent non-spatial problems, including context, probabilities and action-dependent information, in support of "model-based" reinforcement learning to complement learning within the striatum.

## INTRODUCTION

The hippocampus and the striatum support episodic and procedural memory, respectively, with the first more involved in the rapid acquisition of experience about locations and situations, and the second more involved in the acquisition of the stimulus-response associations by means of slower cumulative trial and error learning. This dichotomy has been linked to "model-based" and "model-free" reinforcement learning.

Recent experiments have shown that these areas are also strongly involved in non-spatial tasks, such as multi-step abstract decision problems.

Our hypothesis is that the hippocampus and striatum are part of a complex and versatile machinery that is able to deal with a vast family of problems that can be transformed into Markov Decision Processes, of which the spatial domain is only a subclass.

[1] Institute of Cognitive Neuroscience, University College London.
17 Queen Square, WC1N 3AX London, UK
* Corresponding author. E-mail: f.chersi@ucl.ac.uk
Cite as: Chersi F., Burgess N. (2016). *Hippocampal and striatal involvement in cognitive tasks: a computational model*. Proceedings of the 6th International Conference on Memory, ICOM16, p. 24

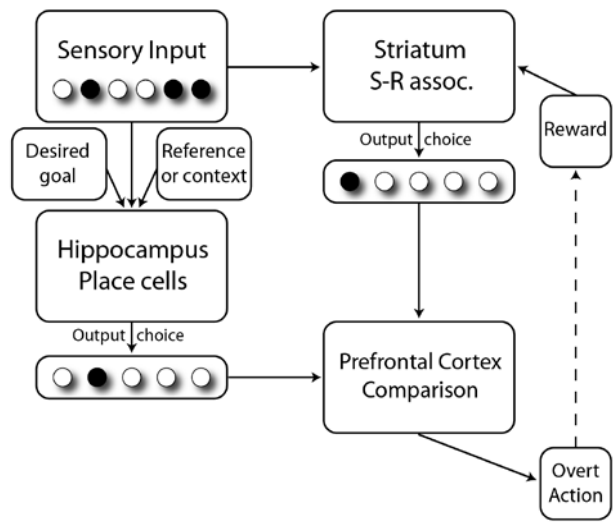

Figure 1. Schematic representation of the hippocampal-striatal circuit for spatial navigation implemented in this work. The common sensory input follows two distinct and parallel paths which at the end converge in the prefrontal cortex to produce an overt action.

## SPATIAL NAVIGATION

A large amount of experimental results have proven that the hippocampus and the striatum play major and mostly complementary roles in spatial navigation.

The hippocampus constitutes the "cognitive map" with specific neurons, called place cells, encoding information about locations (for goal-directed decision making) in an absolute reference frame, while the striatum learns stimulus-response associations (i.e. hardwires specific sensory inputs to corresponding motor outputs) in an egocentric manner. A schematic representation of the architecture developed in this work is represented in Figure 1. Note that the same sensory input is used in different ways by the two systems: the striatum learns only when a reward signal is provided (sensory-motor connections are learned only when the outcome of the action produces a positive or negative result), while the hippocampus uses also additional information about the reference system (e.g. head direction) or context to build a map of objects and locations (thus the goal representation is flexible). Each brain area outputs the estimated optimal action (in our case the turning angle), which is then compared and selected by the prefrontal cortex to produce the corresponding overt action.

As a testbed for our model we replicated the "Plus maze experiment" described by Packard and McGaugh (1996) and shown in figure 2.

The left panel explains how rats are trained to find food from a starting location: animals are always placed at the end of the same arm of the maze, while food is always placed in the same location, then animals are let free to find the food. The right panel describes the behavior of control and treated rats tested for their ability in finding the correct food location when, in unrewarded probe trials, the starting position is moved to the opposite side of the maze, and (see also Chersi and Burgess, 2015).

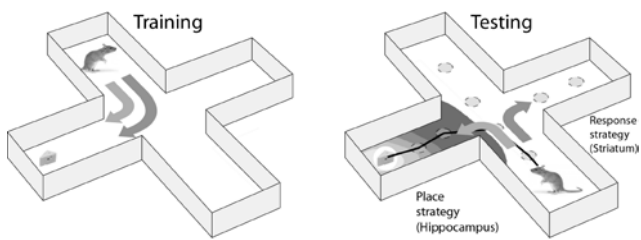

Figure 2. Illustration of the plus maze task used by Packard and McGaugh (1996). Green and orange arrows indicate place and response behavior, respectively. During the learning phase (left) rats are repeatedly trained to move from one arm of the maze to another (always the same). During the unrewarded testing phase (right), depending on the amount of training and the condition of their hippocampus and striatum, placed in a different starting position rats either move to the previously learned location or in the previously learned relative direction.

In this work we implemented a biologically realistic model of the striatal and hippocampal circuits (shown in Figure 1) utilizing firing rate-based neurons endowed with Hebbian and reinforcement learning rules. We also developed a 2D simulator of the environment which allows to obtain simple visual inputs and to control a virtual agent (see also Chersi 2014).

Below we compare the experimental results obtained by Packard and McGaugh (1996) (see Figure 3, left and middle panels), to those we obtained by using the model exemplified in figure 1 and described in Chersi and Burgess (2015) (see Figure 3, right panel). As can be noted initially (Test Day 8, yellow highlighted bar) animals favor a place-based response strategy, while later (Test Day 16, yellow highlighted bar) they switch to response-based strategy. Interestingly our model is capable to easily reproduce these experimental results.

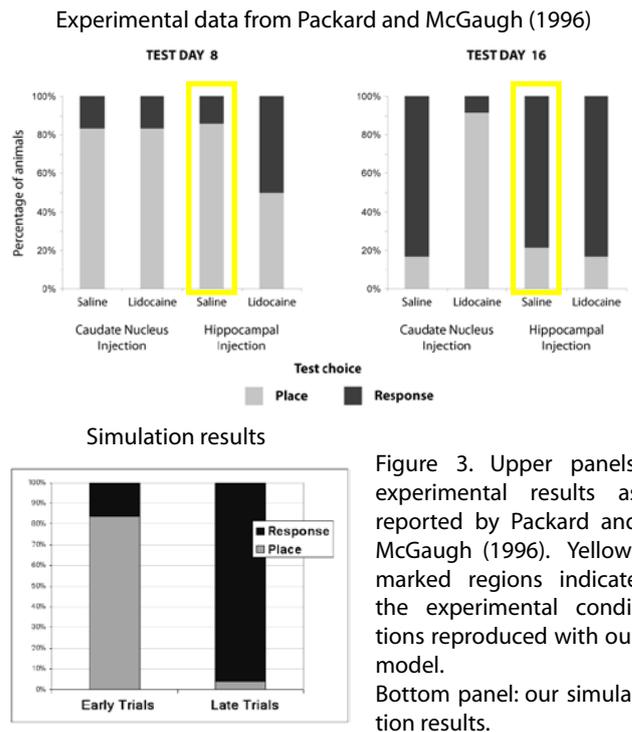

Figure 3. Upper panels: experimental results as reported by Packard and McGaugh (1996). Yellow-marked regions indicate the experimental conditions reproduced with our model.
Bottom panel: our simulation results.

## ABSTRACT DECISION MAKING

In the second part of this work we investigated how the previously described architecture can be utilized to solve more complex types of tasks. In particular we focused on two specific aspects: abstract reasoning and non-deterministic problems.

There are two important questions we tried to answer:

1) How can the hippocampus and striatum solve planning and decision-making problems that do not involve movements in the physical space?

2) How can the brain handle non-deterministic problems?

To answer these questions we took inspiration from the work of Daw et al. (2011).

In their experiments, human subjects were initially presented with two cards of which they had to chose one (see Figure 4). Depending on their choice they were presented with one of two pairs of other cards with a 70%-30% and 30%-70% probability, respectively. Among these they had to chose again one more card. This final choice led to a fixed amount of reward (**R**) but with a slowly varying probability.

It should be noted that, while the rules and the structure of the task is relatively simple, its intrinsic randomness impedes the learning of an optimal habitual choice sequence.

Experimental data has revealed that subjects seem to calculate what they believe to be the best choice sequence by taking into account the task structure that they have learned during the trials and the outcomes of a few previous trials (see Figure 5).

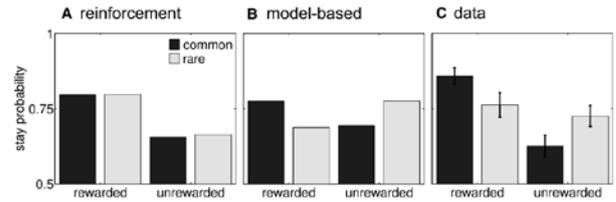

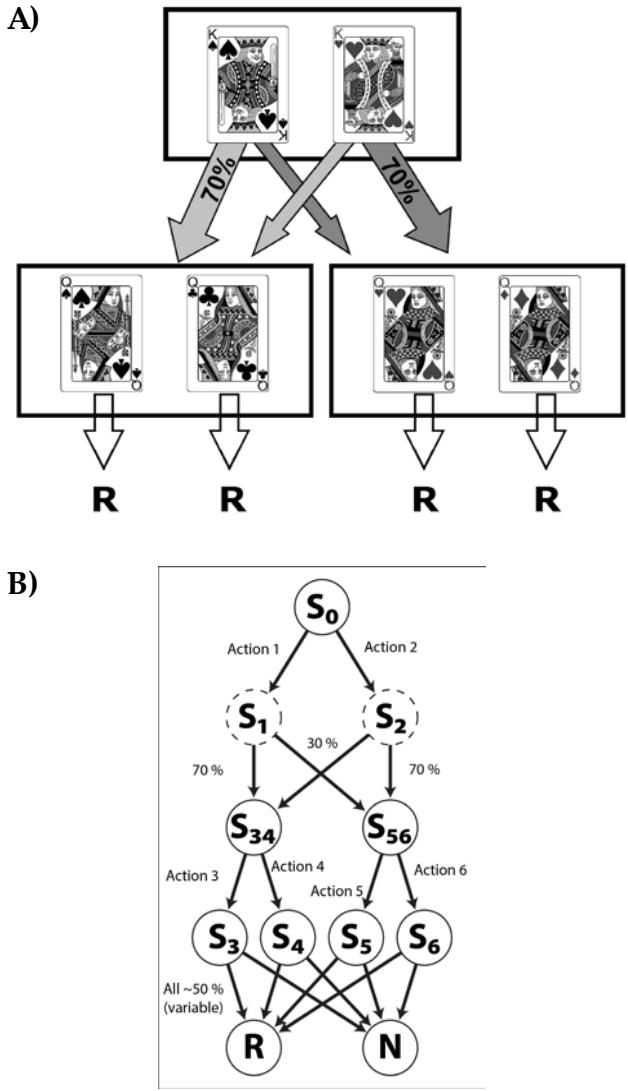

Figure 5. Left and middle panel: predicted probability of choosing the previous action for subjects utilizing only a model-free or a model-based strategy, respectively. Right panel: actual "stay" proportions, averaged across subjects, display hallmarks of both strategies (Daw et al. 2011).

## OUR HYPOTHESES

### Non-spatial problems

According to our hypothesis, the hippocampal and striatal circuits for spatial navigation can be utilized straight away for non-spatial problems if one considers that their representations are in the first place sensorial and not spatial. The fact that sensory cues are mostly associated with locations (at least for animals) has sometimes led to the conception that the hippocampus represents only physical space.

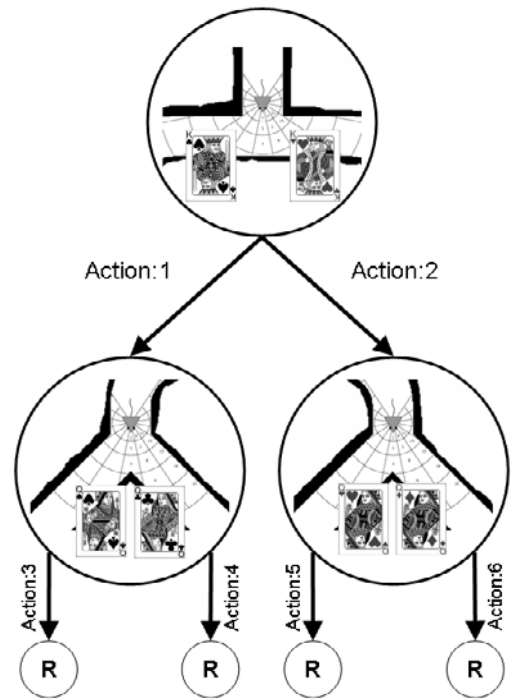

Figure 4. Panel A: experiment by Daw et al. (2011). In this study a two-step choice task is used to investigate the contributions of model-based and model-free mechanisms in decision making. In particular, human subjects are initially presented with two cards of which they have to choose one. Thereafter, they are presented with one of two other sets of cards (with 70% and 30% probability depending to their first choice. At this point they are required to choose one of the two new cards, each one providing a fixed reward with a variable probability (oscillating between 25% and 75%).

Panel B: equivalent Markov Chain. $S_0$ is the initial state when the first two cards are given. As a result of the first choice (Action 1 or 2), one of two pairs of cards ($S_{34}$ or $S_{56}$) are given. After the second choice (Action 3, 4, 5 or 6) a reward **R** or no reward (**N**) might be given with an average probability of 50% (but randomly varying).

Figure 6. A non-spatial problem can be "converted" into a spatial one by imagining to move in an environment were the visual cues are the ones provided in the task at hand. In our case, the cards may be thought of as landmarks in proximity of decision points in a maze, each associated with a specific path.

In order to reconcile these two apparently different types of application domains, we observe that non-spatial problems contain a great amount of sensory information (e.g. in our case the colors and the symbols on the cards). These signals have here being used as an input to the hippocampal and the striatal circuit of Figure 1, exactly in the same way as before when we were using information about the environment (see Figure 6).

More precisely, one can imagine to be in a maze where at crossing points there are the same visual cues that are present in the task at hand (i.e. the cards). On the basis of these cues one will decide to go left or right, which in the current task would be equivalent to choosing the left or the right card. At this point, one can map the whole task structure (Figure 4B) onto an equivalent physical maze representation with decision points and reward locations, and then use the hippocampal and striatal circuits to solve the maze and thus its corresponding counterpart.

As a result we predict that experimental tests should find the same neural activation patterns in the hippocampus and the striatum for spatial and non-spatial tasks that have the same structure.

**Non-deterministic problems**

Physical tasks because of the nature of reality, are necessarily deterministic (except possibly because of erroneous motor execution), i.e. if we move to the right our body will move to the right. On the other hand, in non-spatial tasks, the outcome of our choices can easily be manipulated to produce random results. Never the less our brain is clearly capable of handling these types of problems.

Our hypothesis is that non-deterministic problems are "unwrapped" in the hippocampus to obtain an explicit and exhaustive representation in form of sequences of deterministic sub-problems that can be evaluated like in the normal case (see Figure 7). Our intuition is that this new kind of representation is achieved in the hippocampus through the use of so called "splitter place cells" (Ainge et al., 2007). These cells possess the peculiar property that multiple neurons encode the same location but their activity pattern is modulated by the final goal of the action sequence.

This kind of behavior is the ideal candidate to solve multi-paths graphs such as the one in Figure 7, because neurons with the same goal (i.e. obtaining a certain amount of reward) but encoding different steps of a sequence, would build independent chains of action sequences that allow activity waves ("forward" and "backward sweeps") to travel undisturbed along one path, returning its independent value.

In practice the mechanisms is the following. Once a given task's structure, transition probabilities and reward distributions have been determined through trial and error, and encoded in the hippocampus by means of splitter place cells, this map is used to "probe" the value of single rewarding locations (i.e. the final cards) and to backtrack the sequence necessary to reach the most frequently rewarding one (see Figure 7b).

Simulation results are shown in Figure 8.

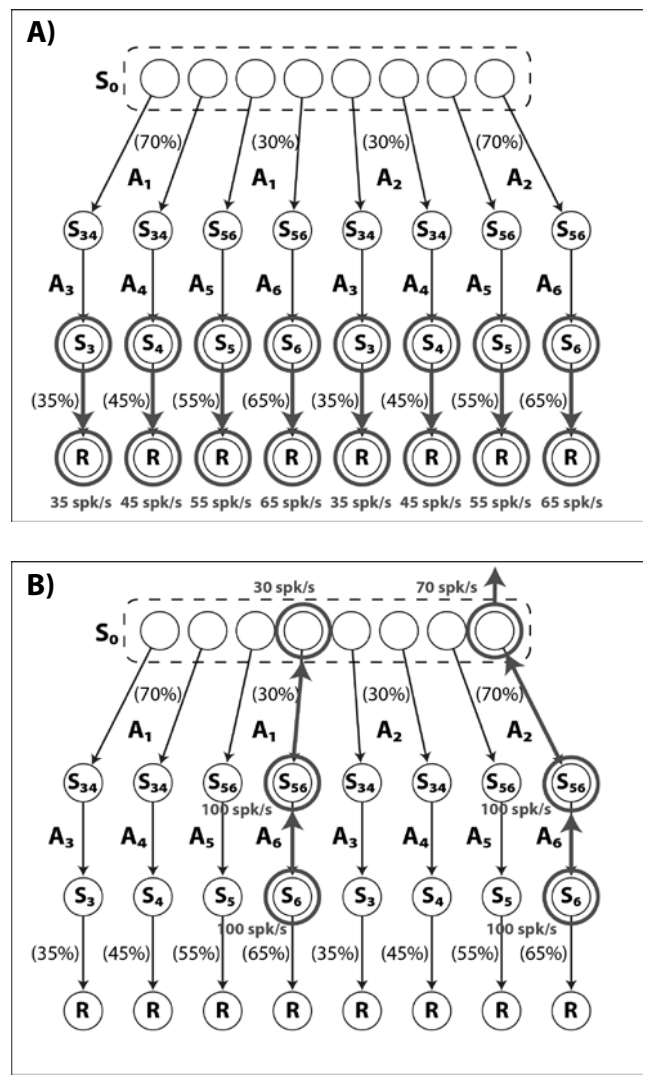

Figure 7. After the structure of the problem, its transition probabilities and its rewards distribution have been learned, and an "unwrapped" representation has been constructed in the hippocampus (panel A), our architecture utilizes a greedy strategy to determine which last card is most frequently rewarding (panel A) and then backtracks the decision tree to find the best strategy that leads to that card (panel B).

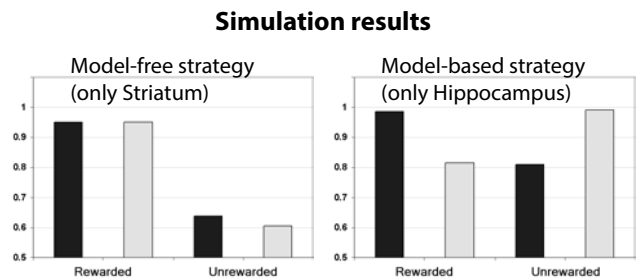

Figure 8. Simulation results utilizing the described model of the hippocampal and striatal circuits. Bars indicate the probability of repeating the previous action. Left panel: only the striatum (trained with a short-memory temporal-difference rule) has been utilized to solve the task. Right panel: only the hippocampus is used to solve the problem. As can be noted, the latter takes into account the structure of the problem, for example by repeating rare (light color) and unrewarded (right column) trials. As can be noted, these results are in very good accord with the data shown in Figure 5.

**CONCLUSIONS**

In this work we have presented a biologically constrained model of the hippocampal and striatal circuits, their functions, and their interactions. These two systems appear to play complementary roles at different stages of spatial learning. The hippocampus learns a goal-independent and allocentric representation of space, thus playing the role of a model-based architecture. The striatum, in contrast, learns egocentric stimulus-response associations, in a model-free manner.

We have made two major contributions with this work:

1) We have shown how the same model can be used to solve spatial and non-spatial decision making tasks.

2) We have shown how the same model can handle deterministic and non-deterministic tasks.

In particular, our first hypothesis is that the brain uses sensory cues present in non-spatial tasks (such as the figures on the cards) to build a map of the abstract sensorial space – complete with transition rules and rewarding states – which can be used to compute and execute sequences of reasoned choices.

Our second hypothesis is that non deterministic tasks (with a limited number of options, i.e. equivalent to Markov Decision Processes) are "unwrapped" in the hippocampus through the use of "splitter place cells" to build equivalent exhaustive deterministic graphs that can be solved in the same way as for deterministic problems.

The type of problems and the solutions addressed in this work have given rise to new and interesting questions for which we provide only speculative answers:

1) How does the hippocampus handle higher dimensional spaces? Since there is no isomorphic mapping in the hippocampus, there is no physical constraint in the encoding of higher dimensional spaces.

2) How are even more abstract kinds of spaces encoded, such as for example tastes, aesthetics or numbers? Since in these cases there are no immediate sensory states, we suppose that the hippocampus can access internally constructed representations and in need build highly abstract cognitive maps.

3) Does the hippocampus represent always the motion of the subject in some type of space? We actually believe that the hippocampus can also represent the motion of single body parts, such as that of the hand or of the eyes, or even of other agents.

4) Can hippocampus-like properties be found in other areas of the brain? It is highly probable that similar properties can be found also in other parts, in particular in the entorhinal cortex.

We are aware that there remain many open questions, but we hope this work has provided insight on some more complex functions of the hippocampal and the striatal circuits, and that it will be of inspiration for future experimental works.

**Bibliography**

Ainge J.A., Tamosiunaite M., Woergoetter F., Dudchenko P.A. (2007). *Hippocampal CA1 place cells encode intended destination on a maze with multiple choice points.* J Neurosci 27, 9769-9779

Chersi F. (2014). *The hippocampal-striatal circuit for goal-directed and habitual choice.* arXiv: 1412.2818

Chersi F., Burgess N. (2015). *The cognitive architecture of spatial navigation: Hippocampal and Striatal contributions.* Neuron 88: 64-77

Daw N., Gershman S., Seymour B., Dayan P., Dolan R. (2011). *Model-Based Influences on Humans' Choices and Striatal Prediction Errors.* Neuron 69, 1205-1215.

Packard M.G., McGaugh J.L. (1996). *Inactivation of hippocampus or caudate nucleus with lidocaine differentially affects expression of place and response learning.* Neurobiol. Learn. Mem. 65 , 65–72.

The research leading to these results has received funding from the European Union Seventh Framework Programme (FP7/2007-2013) under grant agreement no. 604102 (Human Brain Project) and from the Wellcome Trust.